\documentclass[12pt,letterpaper]{JHEP3}

\def\Box{\square}
\def\t#1{\mathrm{#1}}
\def\c#1{\mathcal{#1}}
\def\pd{\partial}

\def\e{\epsilon}
\def\->{\rightarrow}
\def\=={\begin{eqnarray*}}
\def\xx{\end{eqnarray*}}
\def\n==#1{\begin{eqnarray} \label{#1}}
\def\nxx{\end{eqnarray}}

\def\shalf{\textstyle\frac{1}{2}\displaystyle}
\def\sfrac#1#2{\textstyle\frac{#1}{#2}\displaystyle}
\newcommand{\comment}[1]{}
\renewcommand\jhep[3] {\href{http://jhep.sissa.it/stdsearch?paper=#1(#2)#3}
{{\it JHEP\ }{\bf #1} (#2) #3}}

\title{Effect of curvature squared corrections in AdS on the viscosity of the dual gauge theory}

\author{Yevgeny Kats and Pavel Petrov \\
Jefferson Physical Laboratory, Harvard University,\\
Cambridge, MA 02138, U.S.A.}

\abstract{We use the real-time finite-temperature AdS/CFT correspondence to compute the effect of general $R^2$ corrections to the gravitational action in AdS space on the shear viscosity of the dual gauge theory. The $R^2$ terms in AdS$_5$ are determined by the central charges of the CFT. We present an example of a four-dimensional gauge theory in which the conjectured lower bound of $1/4\pi$ on the viscosity-to-entropy ratio is violated for finite $N$.}

\begin{document}


\section{Introduction\label{sec-Intro}}

It has been conjectured~\cite{bound-1,bound-2} that there exists a lower bound $\eta/s \geq 1/4\pi$ on the ratio between the shear viscosity $\eta$ and the entropy density $s$ for a field theory at a finite temperature. Much of the evidence for this conjecture comes from methods based on the AdS/CFT correspondence~\cite{AdS/CFT-Maldacena,AdS/CFT-GKP,AdS/CFT-Witten} that relates a theory of gravity in $D$-dimensional anti-de Sitter (AdS) space to a conformal field theory (CFT) on the $(D-1)$-dimensional boundary. In particular, an AdS space with a black brane is dual to a field theory at a finite temperature~\cite{AdS/CFT-thermal}. The real-time (i.e., Lorentzian signature) finite-temperature AdS/CFT correspondence~\cite{AdS/CFT-RT-prescription,AdS/CFT-RT} allows one to compute various hydrodynamic quantities of the boundary theory in its strongly coupled regime by doing a supergravity calculation in the AdS space. It has been shown, first for particular examples~\cite{1/4pi-1,1/4pi-2,1/4pi-3,bound-1} and later in general~\cite{1/4pi-SG-universal}, that gauge theories that are dual to supergravity theories in AdS (without higher order terms) actually saturate the bound. The bound is still saturated to leading order when fundamental matter is included~\cite{fundamental}. It was also found that in the $\t{AdS}_5\times \t{S}^5$ setup dual to $\c{N}=4$ SYM, the leading $\alpha'$ corrections from string theory (which are quartic in the curvature) increase the ratio $\eta/s$~\cite{IIB-corr}. In the weakly coupled regime, direct calculations using the Boltzmann equation give $\eta/s \gg 1$~\cite{weakly-coupled}. There has been a debate~\cite{bound-1,bound-2,counterexamples,Bekenstein} over whether the bound may apply to all physical fluids, including non-relativistic ones. It is desirable to consider the effect of additional types of corrections to the strongly coupled limit in order to obtain more evidence or a counterexample to the bound conjecture and to gain more insight into the meaning of the ratio $\eta/s$. In addition to these fundamental questions, the results may turn out to be relevant to the quark-gluon plasma, whose viscosity is measured to be close to the bound~\cite{RHIC} (and possibly violates the bound~\cite{RHIC-violates-1,RHIC-violates-2}).

Here we consider the effect of curvature squared corrections, described by the action
\n=={action}
S = \int d^Dx\,\sqrt{-g} \left(\frac{R}{2\kappa} - \Lambda
+ c_1\,R^2
+ c_2\,R_{\mu\nu}R^{\mu\nu}
+ c_3\,R_{\mu\nu\rho\sigma}R^{\mu\nu\rho\sigma}
\right)
\nxx
in $D$ spacetime dimensions, where $c_i$ are arbitrary small coefficients, and the negative cosmological constant $\Lambda$ creates an AdS space with radius
\n=={AdSL}
L^2 = \frac{(D-1)(D-2)}{2\kappa(-\Lambda)}
\nxx
We take a black brane solution and use the real-time AdS/CFT method~\cite{AdS/CFT-RT} and the Kubo formula to compute the viscosity of the dual boundary theory. The corrected viscosity-to-entropy ratio, to first order in $c_i$, is found to be
\n=={ratio-intro}
\frac{\eta}{s} = \frac{1}{4\pi}\left[1 - 4(D-4)(D-1)\frac{c_3}{L^2/\kappa}\right]
\nxx
While $c_1$ and $c_2$ affect the viscosity $\eta$ and the entropy $s$, they do not affect the ratio $\eta/s$, which is indeed expected because of the possibility of a field redefinition, as explained in appendix~\ref{app-field-redef}. More interesting is the non-zero effect on $\eta/s$ due to $c_3$. The conjecture that $1/4\pi$ is the lower bound for $\eta/s$ would require $c_3$ to be \emph{negative}.

It is interesting to ask whether there exist any general consistency conditions on a gravitational theory that fix the sign of the coefficient $c_3$. For example, it is known that many classes of low-energy effective field theories are inconsistent with string theory (and some are inconsistent with \emph{any} theory of quantum gravity)~\cite{swampland,weak-grav,causality,o-vafa-swamp,ebh}. In particular, the coefficients of the low-energy effective action must satisfy certain inequalities because of causality (or analyticity in the UV)~\cite{causality} or the requirement that all charged black holes can decay~\cite{weak-grav,ebh}. However, these are not restrictive enough to fix the sign of $c_3$.

On the other hand, for 4-dimensional CFTs that have AdS$_5$ gravity duals, it was shown~\cite{Blau-Narain-Gava,Nojiri-Odintsov} that in the limit of large $N$ and large 't Hooft coupling $\lambda \equiv g_\t{YM}^2 N = L^4/\alpha'^2$ (while the string coupling $g_s = \lambda/(4\pi N)$ is kept small), $c_3$ is related to the central charges of the CFT as\footnote{The central charges $a$ and $c$ of a four-dimensional CFT are defined in terms of the trace anomaly as $\langle {T^\mu}_\mu\rangle = \frac{1}{16\pi^2}\left(c W - a G\right)$ (when we consider the CFT coupled to an external metric) where $G = R_{\mu\nu\rho\sigma}R^{\mu\nu\rho\sigma} - 4R_{\mu\nu}R^{\mu\nu} + R^2$ is the Gauss-Bonnet combination and $W = R_{\mu\nu\rho\sigma}R^{\mu\nu\rho\sigma} - 2R_{\mu\nu}R^{\mu\nu} + \sfrac{1}{3}R^2$ is the square of the Weyl tensor. Both $a$ and $c$ are $\c{O}(N^2)$ and in order for a supergravity dual to exist they must be equal at leading order in $N$~\cite{Henningson-Skenderis} and given by $a = c = \pi^3N^2/(4V_5)$, where $V_5$ is the volume of the compact manifold (without the $L^5$ factor)~\cite{Gubser}. However, they may differ at subleading order: $c - a \sim \c{O}\left(N\right)$. }
\n=={c3ac}
\frac{c_3}{L^2/\kappa} = \frac{c-a}{16\,c} + {\cal O}\left(1/N^2\right)
\nxx
This gives us
\n=={a/c}
\frac{\eta}{s} = \frac{1}{4\pi}\frac{a}{c} + {\cal O}\left(1/N^2\right)
\nxx
For the well-known $\c{N} = 4$ $\t{SU}(N)$ SYM, $a = c$, so the $c_3$ term does not appear, but in general this does not need to be the case. For example, consider the $\c{N}=2$ $\t{Sp}(N)$ gauge theory with 4 fundamental and 1 antisymmetric traceless hypermultiplets. This superconformal theory arises in string theory in the setup of $N$ D3-branes sitting inside 8 D7-branes coincident on an orientifold 7-plane, and its gravity dual is type IIB string theory on $\t{AdS}_5\times \t{X}^5$, where $\t{X}^5 \simeq \t{S}^5/\mathbb{Z}_2$~\cite{Sp(N),AFM}. The $c_3$ term in this case comes from the effective action on the worldvolume of the D7-branes/O7-plane system~\cite{Aharony,Blau-Narain-Gava}. This theory has $a = \frac{1}{24}(12N^2 + 12N - 1)$, $c = \frac{1}{24}(12N^2 + 18N - 2)$~\cite{Blau-Narain-Gava}, and then for $N\gg1$ eq. (\ref{a/c}) gives
\n=={ratio-example}
\frac{\eta}{s} = \frac{1}{4\pi}\left(1 - \frac{1}{2N}\right)
\nxx
Thus the conjectured bound on $\eta/s$ is violated.

Note that $R^4$ terms from the 10-dimensional bulk contribute an $\c{O}\left(1/\lambda^{3/2}\right)$ correction to $\eta/s$ \cite{IIB-corr}. However, that contribution is suppressed relative to the $\c{O}\left(1/N\right)$ result in (\ref{ratio-example}) in the range $1 \ll \lambda \ll N \ll \lambda^{3/2}$. Contributions from even higher powers of $R$ in the bulk and powers higher than $R^2$ in the brane effective action are suppressed by additional factors of $1/\lambda^{1/2}$. Note also that since the contribution of our $R^2$ term to the action is suppressed by $1/N$, the contribution from any new field that could be sourced as a result (the Kaluza-Klein modes) would be suppressed by $1/N^2$, and thus can be neglected (we keep $g_s$ fixed at a small value and do a large $N$ expansion).
\\\textbf{Note added:} A more detailed justification has been provided later in~\cite{justification}, along with further illuminating discussions.

The structure of the paper is as follows. In section~\ref{sec-metric} we discuss the corrections to the metric of the AdS black brane. Section~\ref{sec-waves} deals with the propagation of waves in the background of the black brane, which then allows us to calculate the viscosity of the boundary theory in section~\ref{sec-viscosity} using AdS/CFT. In section~\ref{sec-ratio} we compute the entropy and find the correction to the viscosity-to-entropy ratio.

\section{AdS black brane metric\label{sec-metric}}

The Einstein-Hilbert action in a $D$-dimensional AdS space
\n=={E-H}
S = \int d^Dx\,\sqrt{-g} \left(\frac{R}{2\kappa} - \Lambda\right)
\nxx
possesses the well-known black brane solution in Poincar\'{e} coordinates
\n=={metric-unperturbed}
ds^2 = \frac{1}{z^2}\left(\frac{-f(z) dt^2 + d\vec x^2}{L^2} + \frac{L^2 dz^2}{f(z)}\right)
\nxx
where
\n=={f(z)}
f(z) = 1 - \left(\frac{z}{z_0}\right)^{D-1}
\nxx
For later convenience, we will change the variable $z$ to $u$ as
\n=={u-defn}
u = \left(\frac{z}{z_0}\right)^\frac{D-1}{2}
\nxx
and then the metric is
\n=={metric-u}
ds^2 = \frac{-f(u)\,dt^2 + d\vec x^2}{L^2z_0^2\,u^\frac{4}{D-1}} + \frac{4L^2}{(D-1)^2}\frac{du^2}{u^2 f(u)} \,,  \quad\quad f(u)= 1-u^2
\nxx

The curvature squared corrections (\ref{action}) modify the function $f(u)$ to
\n=={f(u)}
f(u) = 1 - u^2 + \alpha + \gamma u^4
\nxx
where
\n=={alpha-gamma}
\alpha = \frac{2(D-4)\kappa}{(D-2)L^2}\left[(D-1)(Dc_1 + c_2) + 2 c_3 \right]\,, \quad\quad
\gamma = 2\frac{\kappa}{L^2}(D-3)(D-4) c_3
\nxx
(for details, see appendix~\ref{app-metric}).

In these coordinates, $u = 0$ is the boundary of the AdS space, and the horizon of the black brane is at
\n=={horizon}
u_H \simeq 1 + \frac{\alpha + \gamma}{2}
\nxx
Note that $t$ is not the natural time coordinate in the boundary theory, because for $u \-> 0$ the metric is proportional to
\n=={boundary-metric}
ds^2 = -f(0)\,dt^2 + d\vec x^2
\nxx
rather than the Minkowski metric. Instead, the boundary time coordinate is $t_b = \sqrt{f(0)}\,t$.

The parameter $z_0$ is related to the temperature of the black brane. Consider the near-horizon geometry, and change the coordinate $u$ to $\rho$ as
\n=={rho}
\rho^2 = \frac{16L^2}{(D-1)^2f'(u_H)^2 u_H^2}f(u)
\nxx
Then the $u$-$t$ part of the metric (\ref{metric-u}) takes the Rindler space form
\n=={Rindler-metric}
ds^2 = -(2\pi T)^2\rho^2 dt_b^2 + d\rho^2
\nxx
where
\n=={temperature}
T = \frac{D-1}{8\pi L^2 z_0}\frac{\left|f'(u_H)\right|}{\sqrt{f(0)}}u_H^\frac{D-3}{D-1} = \frac{D-1}{4\pi L^2 z_0}\left[1 + \frac{D-3}{2(D-1)}\alpha - \frac{D}{D-1}\gamma\right]
\nxx
is the temperature.

\section{Waves in the AdS black brane background\label{sec-waves}}

We now add a small fluctuation $\phi(t,u)$:
\n=={fluc-metric}
ds^2 = \frac{-f(u)\,dt^2 + d\vec x^2 + \phi(t,u)\left(dx_1 dx_2 + dx_2 dx_1\right)}{L^2z_0^2\,u^\frac{4}{D-1}} + \frac{4L^2}{(D-1)^2}\frac{du^2}{u^2 f(u)}
\nxx
In the framework of the AdS/CFT correspondence, the field $\phi$ corresponds to the component $T_{12}$ of the energy-momentum tensor of the boundary theory, whose correlators will be used to calculate the viscosity. We take
\n=={FT}
\phi(t,u) = \phi(u)\,e^{-i\omega t_b} = \phi(u)\,e^{-i\sqrt{f(0)}\,\omega t}
\nxx
Treating the corrections to Einstein equations via $T_{\mu\nu}^\t{eff}$ from (\ref{Einstein-corr}), we obtain the equation of motion for $\phi(u)$:\footnote{In (\ref{phi-eqn}), $f(u)$ is just a shorthand for our particular expression (\ref{f(u)}). Also, here and throughout the paper we work to leading order in the coefficients $c_i$.}
\n=={phi-eqn}
\phi'' - \left[\frac{1}{u} + \frac{2u}{f(u)} + \frac{8\gamma}{D-3}\frac{u}{f(u)}\left(1 - \frac{D-1}{2}u^2\right)\right]\phi' + \frac{\bar\omega^2}{f^2(u)\,u^\frac{2(D-3)}{D-1}}\phi = 0
\nxx
where we denoted
\==
\bar\omega \equiv \frac{2\sqrt{f(0)}L^2z_0}{D-1}\,\omega
\xx
To solve (\ref{phi-eqn}), we first consider the singular behavior at the horizon where $f(u) = 0$. There the equation reduces to
\n=={phi-eqn-horizon}
\phi'' - \frac{2}{f(u)}\left(1 + \frac{\alpha-3\gamma}{2}\right)\phi' + \frac{\bar\omega^2}{f^2(u)}\left(1 - \frac{D-3}{D-1}(\alpha+\gamma)\right)\phi = 0
\nxx
Writing $\phi(u) = f(u)^\nu F(u)$ with $F(u)$ regular at the horizon, and leaving only the most divergent terms, we obtain two possible values
\n=={nu}
\nu_\pm = \pm i\frac{\bar\omega}{2}\left(1 - \frac{D-2}{D-1}\alpha + \frac{D}{D-1}\gamma\right) = \pm i\frac{\omega}{4\pi T}
\nxx
The full eq. (\ref{phi-eqn}) then gives the following equation for $F(u)$:
\n=={F-eqn}
&&F'' - \left[\left((1+2\nu)\left(1 - \frac{\alpha + \gamma u^4}{1-u^2}\right) + \frac{4\gamma}{D-3}\right)\frac{2u}{1-u^2} - 4\left(2\nu + \frac{D-1}{D-3}\right)\gamma\frac{u^3}{1-u^2} + \frac{1}{u}\right]F' \nonumber\\
&&\quad\quad + 8\nu\frac{D-1}{D-3}\frac{u^2}{1-u^2}\gamma F = 0
\nxx
We dropped $\bar\omega^2$ terms, since it will be sufficient for us to keep in $\phi(u)$ only terms that are linear in $\omega$. If we ignore terms proportional to $c_i$, we have a solution
\n=={F-unperturbed}
F(u) = 1
\nxx
Substituting this solution into terms that are proportional to $c_i$ in (\ref{F-eqn}), we are left with
\n=={F-eqn-2}
F'' - \left[(1+2\nu)\frac{2u}{1-u^2} + \frac{1}{u}\right]F' + 8\nu\frac{D-1}{D-3}\frac{u^2}{1-u^2}\gamma = 0
\nxx
The solution is
\n=={F-gen-soln}
F(u) &=& k_1 + k_2 (1 - u^2)^{-2\nu} + 2\nu\gamma\frac{D-1}{D-3}u^2
\nxx
where $k_1$ and $k_2$ are integration constants. We take $k_1 = 1$, consistently with our unperturbed solution $F(u) = 1$, and we must take $k_2 = 0$ to have $F(u)$ regular at the horizon, so
\n=={F-soln}
F(u) = 1 + 2\nu\gamma\frac{D-1}{D-3}u^2
\nxx
Then the full $\phi(u)$, to first order in $\omega$, and normalized to $\phi(0) = 1$, is
\n=={phi-pm}
\phi_\pm(u) = 1 \pm i\frac{\omega}{4\pi T}\left[\ln(1-u^2) + \frac{\alpha + \gamma u^4}{1-u^2} - \alpha + 2\gamma\frac{D-1}{D-3}u^2\right]
\nxx
and the general solution can be written as
\n=={phi-gen}
\phi(u) = a\,\phi_+(u) + b\,\phi_-(u)
\nxx
where $a$ and $b$ are constants.

\section{Viscosity via AdS/CFT\label{sec-viscosity}}

We will calculate the shear viscosity of the boundary theory using the Kubo formula:
\n=={Kubo}
\eta = \lim_{\omega\->0}\frac{1}{2\omega}\int dt d\vec x\,e^{i\omega t}\,\langle[T_{12}(x),T_{12}(0)]\rangle = -\lim_{\omega\->0}\frac{1}{\omega}\,\t{Im}\,G^R(\omega,0)
\nxx
where $G^R$ is the retarded Green's function for $T_{12}$:
\==
G^R(\omega, \vec q) = -i\int dtd\vec x\,e^{-i\vec q\cdot\vec x + i\omega t}\,\theta(t)\,\langle[T_{12}(x),T_{12}(0)]\rangle
\xx

To find $G^R(\omega,0)$ we will use the AdS/CFT correspondence, where the bulk field corresponding to the boundary operator $T_{12}$ is our field $\phi(t,u)$. In Euclidean space, the correspondence says~\cite{AdS/CFT-Witten}
\n=={AdS/CFT}
\left\langle \exp\left(\int d^{D-1}x\,\phi_0(x)\,T_{12}(x)\right)\right\rangle_\t{CFT} = \left.e^{-S_\t{SG}[\phi]}\right|_\t{AdS}
\nxx
Here $\phi$ is the classical solution to the supergravity equations of motion, subject to the condition that the boundary value of $\phi$ is $\phi_0(x)$, and $S_\t{SG}[\phi]$ is the corresponding value of the supergravity action. Taking two functional derivatives with respect to $\phi_0$ and setting $\phi_0 = 0$, we see that the part of $S_\t{SG}$ that is quadratic in $\phi_0$ can be written in terms of the Euclidean Green's function $G(x-x') \equiv \langle T_{12}(x)T_{12}(x')\rangle$ as
\n=={AdS/CFT-Green}
S_\t{SG}^{(2)} = -\frac{1}{2}\int\frac{d^{D-1}q}{(2\pi)^{D-1}}\,\phi_0(q)\,G(q)\,\phi_0(-q)
\nxx
where we switched to momentum space. The possibility to adapt this relation to the case of a Lorentzian theory at a finite temperature has been discussed by Herzog and Son~\cite{AdS/CFT-RT}. According to their argument, one needs to consider the fully extended Penrose diagram for the AdS black brane and set the boundary condition that the solution includes only positive-frequency modes leaving into the future horizon and negative-frequency modes emerging from the past horizon, where the notion of positive/negative frequency should be defined with respect to Kruskal coordinates. The analog of (\ref{AdS/CFT-Green}) is
\n=={Schwinger-Keldysh}
S_\t{SG}^{(2)} = \frac{1}{2}\int\frac{d^{D-1}q}{(2\pi)^{D-1}}\,\phi_i(q)\,G_{ij}(q)\,\phi_j(-q) \nxx
where $G_{ij}(q)$ (with $i,j=1,2$) are the Schwinger-Keldysh propagators for the operators of the thermal theory and their doubler operators (see ref.~\cite{AdS/CFT-RT} for a review), and $\phi_1$ and $\phi_2$ are sources for these fields. In the AdS/CFT setup, $\phi_1$ corresponds to the boundary value of our field $\phi$, while $\phi_2$ corresponds to the boundary value of a similar field that lives on the left Rindler-like patch of spacetime, whose behavior is also described by (\ref{phi-pm}). We will call these fields $\phi_R$ and $\phi_L$, respectively. Unruh~\cite{Unruh} has shown that in order to construct purely positive or negative frequency modes one needs to take the following combinations:
\n=={Unruh-combinations}
\phi_\t{neg} = \left\{\begin{array}{cc}
\phi_+ & \quad\t{Right} \\\\
e^{\omega/2T}\phi_+ & \quad\t{Left}
\end{array}\right.
\qquad\qquad
\phi_\t{pos} = \left\{\begin{array}{cc}
\phi_- & \quad\t{Right} \\\\
e^{-\omega/2T}\phi_- & \quad\t{Left}
\end{array}\right.
\nxx
This condition relates the coefficients $a$ and $b$ from (\ref{phi-gen}) of the left patch to those of the right patch. Further, we can set the boundary condition $\phi_L = 0$ on the AdS boundary of the left patch. Then the full solution in the right and the left patches is given by
\n=={phi}
\phi_R(u) &=& a(\phi_+(u) - e^{\omega/T}\phi_-(u)) \nonumber\\
&=& 1 - \frac{i}{2\pi}\left[\ln(1-u^2) + \frac{\alpha + \gamma u^4}{1-u^2} - \alpha + 2\gamma\frac{D-1}{D-3}u^2\right] + \c{O}(\omega) \\
\phi_L(u) &=& a\,e^{\omega/2T}\left[\phi_+(u) - \phi_-(u) \right] \nonumber\\
&=& -\frac{i}{2\pi} \left[\ln(1-u^2) + \frac{\alpha + \gamma u^4}{1-u^2} - \alpha + 2\gamma\frac{D-1}{D-3}u^2\right] + \c{O}(\omega)
\nxx
where in the last step we picked a convenient value for $a$. Eq. (\ref{Schwinger-Keldysh}) becomes
\n=={action-G11}
S_\t{SG}^{(2)} = \left.\frac{V_{D-2}}{2}\int\frac{d\omega}{2\pi}\;\phi_{R,\omega}(u)\,G_{11}(\omega)\,\phi_{R,-\omega}(u)\right|_\t{boundary}
\nxx

Our supergravity action, keeping only terms quadratic in $\phi$, is given by
\n=={action-phi}
S_\t{SG}^{(2)} &=& \int du\,d^{D-1}x\,\sqrt{-g}\left(\frac{R}{2\kappa} - \Lambda + c_1 R^2 + c_2 R_{\mu\nu}R^{\mu\nu} + c_3\,R_{\mu\nu\rho\sigma}R^{\mu\nu\rho\sigma}\right) \nonumber\\
&=& \frac{V_{D-2}}{16\kappa L^D z_0^{D-1}}\int du\,dt\,[A(u)\phi\phi'' + B(u)\phi'^2 + C(u)\phi\phi' \nonumber\\
&&\qquad\qquad\qquad\qquad\quad + D(u)\phi^2 + E(u)\phi''^2 + F(u)\phi'\phi'']
\nxx
where $V_{D-2} = \int d^{D-2}x$, and the functions $A(u)$, $B(u)$, etc. are given in appendix~\ref{app-action-coeffs}. Identically to ref.~\cite{IIB-corr}, we add the appropriate Gibbons-Hawking boundary terms, and obtain the total action as the boundary term
\n=={action-mom-space}
S_\t{SG}^{(2)} = \frac{V_{D-2}}{16\kappa L^Dz_0^{D-1}\sqrt{f(0)}} \left.\int \frac{d\omega}{2\pi}\c{F}_\omega(u)\right|_\t{boundary}
\nxx
where
\n=={F_q}
\c{F}_\omega(u) &=& \frac{1}{2}(C-A')\phi_\omega\phi_{-\omega} + \left(B - A - \frac{F'}{2}\right)\phi'_\omega\phi_{-\omega} + \nonumber\\
&& + E(\phi''_\omega\phi'_{-\omega} - \phi'''_\omega\phi_{-\omega}) - E'\phi''_\omega\phi_{-\omega} - E\frac{1+u^2}{u(1-u^2)}\phi'_\omega\phi'_{-\omega}
\nxx
Here $\phi_\omega(u)$ is the Fourier transform of $\phi(t,u)$ with respect to $t_b$, as in (\ref{FT}); $\c{F}_\omega(u)$ should be evaluated for both $\phi_R$ and $\phi_L$  (however, the latter happens not to contribute to the final result for $\eta$). Based on (\ref{action-G11}), we can read off $G_{11}(\omega)$ from (\ref{action-mom-space}):
\n=={G11}
G_{11}(\omega) = \left.\frac{1}{8\kappa L^D z_0^{D-1}\sqrt{f(0)}}\frac{\c{F}_\omega(u)}{\phi_{R,\omega}(u)\phi_{R,-\omega}(u)}\right|_\t{boundary}
\nxx
Using the general relation
\n=={G11-GR}
G_{11}(q) = \t{Re}\,G^R(q) + i\coth\frac{\omega}{2T}\,\t{Im}\,G^R(q)
\nxx
in the limit $\omega \-> 0$, eq. (\ref{Kubo}) gives the viscosity as
\n=={visc-F}
\eta = -\frac{1}{2T}\,\t{Im}\,G_{11}(0)
= -\left.\frac{1}{16\kappa L^D z_0^{D-1}\sqrt{f(0)}\,T}\,\t{Im}\frac{\c{F}_0(u)}{\phi_{R,0}^2(u)}\right|_\t{boundary}
\nxx
We then obtain\footnote{Technically, only the term $\left(B - A\right)\phi'_\omega\phi_{-\omega}$ from (\ref{F_q}) contributes to the unperturbed $\eta$, while the $c_i$ corrections receive a contribution from $E'\phi''_\omega\phi_{-\omega}$ as well.}
\n=={viscosity}
\eta = \frac{1}{2\kappa (Lz_0)^{D-2}}\left[1 - \frac{6(D-2)(Dc_1 + c_2)}{L^2/\kappa} - \frac{2(D-4)(D^2-5D+8)}{D-1}\frac{c_3}{L^2/\kappa}\right]
\nxx

\section{Viscosity-to-entropy ratio\label{sec-ratio}}

We calculate the entropy of the black brane (which is also the entropy of the boundary theory) using Wald's formula~\cite{Wald-1,Wald-2,Jacobson-Kang-Myers,Brustein-Gorbonos-Hadad}. We have the action $S = \int d^Dx\,\sqrt{-g}\,L$ with
\n=={L}
L(g_{\mu\nu},R_{\mu\nu\rho\sigma}) &=& \frac{R}{2\kappa} - \Lambda + c_1 R^2 + c_2 R_{\mu\nu}R^{\mu\nu} + c_3 R_{\mu\nu\rho\sigma}R^{\mu\nu\rho\sigma}
\nxx
and according to Wald's formula, the entropy $\c{S}$ is given by
\n=={Wald}
\c{S} = -2\pi\oint d^{D-2}x\,\sqrt{h}\,\frac{\pd L}{\pd R_{abcd}}\e_{ab}\e_{cd}
\nxx
where the integral is over the surface of the horizon, $h$ is the determinant of the induced metric, and $\e_{ab}$ is the binormal normalized as $\e_{ab}\e^{ab} = -2$. Then for our metric (\ref{metric-u}), the entropy per unit $(D-2)$-dimensional volume of the boundary theory is given by
\n=={entropy-1}
s &=& \left. -2\pi\sqrt{h}\,\frac{\pd L}{\pd R_{abcd}}\e_{ab}\e_{cd}\right|_\t{horizon} \nonumber\\\nonumber\\
&=& 2\pi\sqrt{h}\left[\frac{1}{\kappa} - \frac{4(D-1)(Dc_1 + c_2 - (D-4)c_3)}{L^2}\right]
\nxx
where the area factor
\n=={h}
\sqrt{h} = \frac{1}{(Lz_0)^{D-2}\,u^\frac{2(D-2)}{D-1}}
\nxx
should be evaluated at the horizon (\ref{horizon}). We then obtain
\n=={entropy}
s = \frac{2\pi}{(Lz_0)^{D-2}}\left[\frac{1}{\kappa} - 6(D-2)\frac{Dc_1 + c_2}{L^2} + \frac{2(D-4)(D-2)(D+3)}{D-1}\frac{c_3}{L^2}\right]
\nxx
The resulting viscosity-to-entropy ratio is
\n=={ratio}
\frac{\eta}{s} = \frac{1}{4\pi}\left[1 - 4(D-4)(D-1)\frac{c_3}{L^2/\kappa}\right]
\nxx
The implications of this result were discussed in the Introduction.

\acknowledgments

\vspace{-3mm}

We are grateful to Nima Arkani-Hamed for suggesting and supervising this project. We thank Ofer Aharony, Vyacheslav Lysov and Andrei Starinets for their enthusiastic responses to our questions, and Hovhannes Grigoryan, Subhaneil Lahiri, Yuji Tachikawa, and especially Juan Maldacena for useful comments on the manuscript. We appreciate communications with Steve Shenker and collaborators who are about to report similar results~\cite{Shenker}.

\appendix

\section{Field redefinitions\label{app-field-redef}}

We start with
\==
S &=& \int d^Dx\,\sqrt{-g}\left(\frac{R}{2\kappa} - \Lambda + c_1 R^2 + c_2 R_{\mu\nu}R^{\mu\nu}\right)
\xx
and consider
\n=={field-redef}
g_{\mu\nu} \-> g_{\mu\nu} + a\,g_{\mu\nu}R + b\,R_{\mu\nu}
\nxx
where $a$ and $b$ are $\c{O}(c_i)$. We then get
\==
S &=& \int d^Dx\,\sqrt{-g}\left(\frac{R}{2\tilde\kappa} - \Lambda + \tilde c_1 R^2 + \tilde c_2 R_{\mu\nu}R^{\mu\nu}\right)
\xx
with
\==
\frac{1}{\tilde\kappa} = \frac{1}{\kappa} - (Da + b)\Lambda \qquad\qquad
\tilde c_1 = c_1 + \frac{(D-2)a + b}{4\kappa} \qquad\qquad
\tilde c_2 = c_2 - \frac{b}{2\kappa}
\xx
In particular, we can eliminate $c_1$ and $c_2$ altogether by taking
\n=={a-and-b}
a = -\frac{2\kappa}{D-2}(2c_1 + c_2)\,,
\qquad\qquad
b = 2\kappa c_2
\nxx
Thus, the action describing fluctuations of $\phi$ in the black brane geometry can be replaced with an action without $c_1$ and $c_2$ but with a different $\kappa$. Since in a theory without $c_1$ and $c_2$ the ratio $\eta/s = 1/(4\pi)$ does not depend on $\kappa$ or anything else, the $c_1$ and $c_2$ terms cannot modify this ratio.  (On the other hand, the viscosity and entropy separately do depend on $\kappa$ and the parameters of the metric, so they can be modified by $c_1$ and $c_2$.)

\section{Computing the corrected black brane metric\label{app-metric}}

Consider an action of the form
\==
S = \int d^Dx\,\sqrt{-g} \left[\frac{R}{2\kappa} - \Lambda
+ c_1\,R^2
+ c_2\,R_{\mu\nu}R^{\mu\nu}
+ c_3\,R_{\mu\nu\rho\sigma}R^{\mu\nu\rho\sigma}
+ \c{L}_m
\right]
\xx
where $\c{L}_m$ is a placeholder for the matter Lagrangian. Varying the action we get
\==
\delta S &=& \int d^Dx\,\sqrt{-g} \left[\frac{1}{2\kappa} \left( R_{\mu\nu} - \shalf R g_{\mu\nu} + \kappa\Lambda g_{\mu\nu}\right)
+ c_1 (\ldots) + c_2 (\ldots) + c_3 (\ldots) - \shalf T_{\mu\nu} \right] \delta g^{\mu\nu} \\
&\equiv& \int d^Dx\,\sqrt{-g} \left[\frac{1}{2\kappa} \left( R_{\mu\nu} - \shalf R g_{\mu\nu} + \kappa\Lambda g_{\mu\nu}\right) - \shalf T_{\mu\nu}^\t{eff} \right] \delta g^{\mu\nu}
\xx
where in the last step we absorbed all the $c_i$ terms in an effective $T_{\mu\nu}$
\n=={Einstein-corr}
T_{\mu\nu}^\t{eff} &=& c_1 \left(g_{\mu\nu} R^2 - 4RR_{\mu\nu} +
4\nabla_\nu\nabla_\mu R - 4g_{\mu\nu}\Box R \right) + \nonumber \\ \nonumber
&+& c_2 \left(g_{\mu\nu} R_{\rho\sigma} R^{\rho\sigma} +
4\nabla_\alpha\nabla_\nu R^\alpha_\mu - 2\Box R_{\mu\nu} -
g_{\mu\nu} \Box R - 4R^\alpha_\mu R_{\alpha\nu} \right) + \\
&+& c_3 \left(g_{\mu\nu}R_{\alpha\beta\gamma\delta}R^{\alpha\beta\gamma\delta} - 4R_{\mu\alpha\beta\gamma}{R_\nu}^{\alpha\beta\gamma} - 8\Box
R_{\mu\nu} + 4\nabla_\nu\nabla_\mu R + 8R_\mu^\alpha R_{\alpha\nu}
- 8R^{\alpha\beta}R_{\mu\alpha\nu\beta} \right) \nonumber \\
\nxx
Since all the terms in $T_{\mu\nu}^\t{eff}$ are already explicitly proportional to $c_i$, we can simply substitute in them the unperturbed solution (\ref{metric-unperturbed}) and treat $T_{\mu\nu}^\t{eff}$ like a non-gravitational source. The calculation of the resulting corrections to the metric is straightforward (similar to ref.~\cite{RN-corr}). One assumes an ansatz of the form
\==
ds^2 = \frac{1}{z^2}\left(-e^{2 a(z)} dt^2 + d\vec x^2 + e^{-2 b(z)}dz^2\right)
\xx
and notes that the components of the Ricci tensor
\==
R^t_t &=& \left[-(D-1) + (D-1)za' + zb' - z^2(a'^2 + a'b' + a'')\right]e^{2b} \\
R^x_x &=& \left[-(D-1) + z(a'+b')\right]e^{2b} \\
R^z_z &=& \left[-(D-1) + za' + (D-1)zb' - z^2(a'^2 + a'b' + a'')\right]e^{2b}
\xx
can be combined as
\n=={Rtt-Rzz}
R^t_t - R^z_z = (D-2)z(a'-b')e^{2b}
\nxx
\n=={Rtt-Rzz-Rxx}
\frac{R^t_t - R^z_z}{D-2} - R^x_x = \left[(D-1) - 2b'z\right]e^{2b} = -\left(\frac{e^{2b}}{z^{D-1}}\right)'z^D
\nxx
Using (\ref{Rtt-Rzz-Rxx}) we can solve for $b(z)$ as
\n=={b-soln}
e^{2b(z)} &=& -z^{D-1}\left[\int\frac{dz}{z^D}\left(\frac{R^t_t - R^z_z}{D-2} - R^x_x\right) + \t{const} \right] \nonumber \\\nonumber\\
&=& -\frac{2\kappa}{D-2}z^{D-1}\left[\int\frac{dz}{z^D}\left(T^t_t - \Lambda\right) + \t{const} \right] \nonumber \\\nonumber\\
&=& \frac{1}{L^2}\left[1 - \left(\frac{z}{z_0}\right)^{D-1}\right] - \frac{2\kappa}{D-2}\,z^{D-1}\int\frac{dz}{z^D}\,T^t_t
\nxx
where we used Einstein's equation in the form
\==
R_{\mu\nu} = \frac{2\kappa}{D-2}g_{\mu\nu}\Lambda + \kappa \left( T_{\mu\nu} - \frac{T}{D-2}\,g_{\mu\nu} \right)
\xx
After found $b(z)$, we can obtain $a(z)$ using (\ref{Rtt-Rzz}):
\n=={a-soln}
a(z) = b(z) + \frac{\kappa}{D-2}\int\frac{dz}{z}e^{-2b(z)}\left(T^t_t - T^z_z\right)
\nxx
By computing (\ref{b-soln}) and (\ref{a-soln}) with the unperturbed solution (\ref{metric-unperturbed}) substituted in (\ref{Einstein-corr}), and using the unperturbed $e^{-2b(z)}$ in (\ref{a-soln}), we find the perturbed metric
\n=={metric-z}
ds^2 = \frac{1}{z^2}\left(\frac{-f(z) dt^2 + d\vec x^2}{L^2} + \frac{L^2 dz^2}{f(z)}\right)
\nxx
where
\n=={f(z)-app}
f(z) = 1 - \left(\frac{z}{z_0}\right)^{D-1} + \alpha + \gamma \left(\frac{z}{z_0}\right)^{2(D-1)}
\nxx
and we defined the constants
\==
\alpha = \frac{2(D-4)\kappa}{(D-2)L^2}\left[(D-1)(Dc_1 + c_2) + 2 c_3 \right]\,, \quad\quad
\gamma = 2\frac{\kappa}{L^2}(D-3)(D-4) c_3
\xx
If we change the variable $z$ to $u$ as
\==
u = \left(\frac{z}{z_0}\right)^\frac{D-1}{2}
\xx
we obtain (\ref{f(u)}).

\section{Coefficients in the action\label{app-action-coeffs}}

The coefficients in the action (\ref{action-phi}) are given, up to terms $\c{O}\left(\omega^2\right)$, by
\==
A(u) &=& 16(D-1)\left[\frac{1-u^2}{2u} + \left(2u - \frac{D}{(D-2)\,u}\right)(D-1)\frac{Dc_1 + c_2}{L^2/\kappa} + \right.\\
&&\qquad\qquad\quad \left. + \left((D-3)(D-2) u^3 - 2(D-5)u - \frac{2D}{(D-2)\,u}\right)\frac{c_3}{L^2/\kappa}\right] \\
B(u) &=& (D-1)\left[6\frac{1-u^2}{u} - \frac{D-1}{(D-2)\,u}\frac{12D^2 c_1 - (D^2-15D+2)c_2}{L^2/\kappa} + \right. \\
&&\left.\qquad\quad + (D-1) u \frac{24Dc_1 + 2(D+11)c_2}{L^2/\kappa} + (D-1)^2 u^3\frac{c_2}{L^2/\kappa} +\right. \\
&&\qquad\quad\left. + \left(\frac{4(D^3-8D^2+19D-26)}{(D-2)\,u} - 16(D^2-6D+3)u + 4(8D^2-43D+49)u^3\right)\frac{c_3}{L^2/\kappa}\right] \\
C(u) &=& -8\left(\frac{D+1}{u^2} + (D-3)\right) + 16(D-1)\left[\frac{D(D+1)}{(D-2)\,u^2} + 2(D-3)\right]\frac{Dc_1 + c_2}{L^2/\kappa} + \\
&&\; + 16\left[\frac{2D(D+1)}{(D-2)\,u^2} - 2(D-3)(D-5) + (D-3)(D-2)(3D-5)u^2\right]\frac{c_3}{L^2/\kappa} \\
D(u) &=& \frac{16}{u^3}\left[1 -\frac{2}{D-2}\frac{D(D-1)(Dc_1 + c_2) + (2D + (D-3)(D-2)^2u^4))c_3}{L^2/\kappa}\right] \\
E(u) &=& (D-1)^3(1-u^2)^2u\frac{c_2 + 4c_3}{L^2/\kappa} \\
F(u) &=& -2(D-1)^2(1-u^2)\frac{(D-1)(1+u^2)c_2 + 4(2(D-3)u^2 - (D-5))c_3}{L^2/\kappa}
\xx



\newpage

\begin{thebibliography}{19}        

\bibitem{bound-1}
P.~Kovtun, D.~T.~Son and A.~O.~Starinets,
\emph{Holography and hydrodynamics: Diffusion on stretched horizons,}
\jhep{10}{2003}{064}
[\hepth{0309213}].

\bibitem{bound-2}
P.~Kovtun, D.~T.~Son and A.~O.~Starinets,
\emph{Viscosity in strongly interacting quantum field theories from black hole physics,}
\prl{94}{2005}{111601}
[\hepth{0405231}].

\bibitem{AdS/CFT-Maldacena}
J.~M.~Maldacena,
\emph{The large $N$ limit of superconformal field theories and supergravity,}
\atmp{2}{1998}{231}
[\ijtp{38}{1999}{1113}]
[\hepth{9711200}].

\bibitem{AdS/CFT-GKP}
S.~S.~Gubser, I.~R.~Klebanov and A.~M.~Polyakov,
\emph{Gauge theory correlators from non-critical string theory,}
\plb{428}{1998}{105}
[\hepth{9802109}].

\bibitem{AdS/CFT-Witten}
E.~Witten,
\emph{Anti-de Sitter space and holography,}
\atmp{2}{1998}{253}
[\hepth{9802150}].

\bibitem{AdS/CFT-thermal}
E.~Witten,
\emph{Anti-de Sitter space, thermal phase transition, and confinement in  gauge theories,}
\atmp{2}{1998}{505}
[\hepth{9803131}].

\bibitem{AdS/CFT-RT-prescription}
D.~T.~Son and A.~O.~Starinets,
\emph{Minkowski-space correlators in AdS/CFT correspondence: recipe and applications,}
\jhep{09}{2002}{042}
[\hepth{0205051}].

\bibitem{AdS/CFT-RT}
C.~P.~Herzog and D.~T.~Son,
\emph{Schwinger-Keldysh propagators from AdS/CFT correspondence,}
\jhep{03}{2003}{046}
[\hepth{0212072}].

\bibitem{1/4pi-1}
G.~Policastro, D.~T.~Son and A.~O.~Starinets,
\emph{Shear viscosity of strongly coupled $\c{N}=4$ supersymmetric Yang-Mills plasma,}
\prl{87}{2001}{081601}
[\hepth{0104066}].

\bibitem{1/4pi-2}
G.~Policastro, D.~T.~Son and A.~O.~Starinets,
\emph{From AdS/CFT correspondence to hydrodynamics,}
\jhep{09}{2002}{043}
[\hepth{0205052}].

\bibitem{1/4pi-3}
C.~P.~Herzog,
\emph{The hydrodynamics of M-theory,}
\jhep{12}{2002}{026}
[\hepth{0210126}].

\bibitem{1/4pi-SG-universal}
A.~Buchel and J.~T.~Liu,
\emph{Universality of the shear viscosity in supergravity,}
\prl{93}{2004}{090602}
[\hepth{0311175}].

\bibitem{fundamental}
D.~Mateos, R.~C.~Myers and R.~M.~Thomson,
\emph{Holographic viscosity of fundamental matter,}
\prl{98}{2007}{101601}
[\hepth{0610184}].

\bibitem{IIB-corr}
A.~Buchel, J.~T.~Liu and A.~O.~Starinets,
\emph{Coupling constant dependence of the shear viscosity in $\c{N}=4$ supersymmetric Yang-Mills theory,}
\npb{707}{2005}{56}
[\hepth{0406264}].

\bibitem{weakly-coupled}
P.~Arnold, G.~D.~Moore and L.~G.~Yaffe,
\emph{Transport coefficients in high temperature gauge theories (I):  leading-log results,}
\jhep{11}{2000}{001}
[\hepph{0010177}].

\bibitem{counterexamples}
T.~D.~Cohen,
\emph{Is there a `most perfect fluid' consistent with quantum field theory?,}
\prl{99}{2007}{021602}
[\hepth{0702136}];
\\
A.~Cherman, T.~D.~Cohen and P.~M.~Hohler,
\emph{A sticky business: the status of the cojectured viscosity/entropy density bound,''}
\jhep{02}{2008}{026}
[\arXivid{0708.4201} [hep-th]];
\\
D.~T.~Son,
\emph{Comment on `Is there a `most perfect fluid' consistent with quantum field theory?',}
\prl{100}{2008}{029101}
[\arXivid{0709.4651} [hep-th]];
\\
T.~D.~Cohen,
\emph{Response to D.T. Son's comment on ``Is there a `most perfect fluid' consistent with quantum field theory?'',}
\arXivid{0711.2664} [hep-th].

\bibitem{Bekenstein}
I.~Fouxon, G.~Betschart and J.~D.~Bekenstein,
\emph{The bound on viscosity and the generalized second law of thermodynamics,}
\prd{77}{2008}{024016}
[\arXivid{0710.1429} [gr-qc]].

\bibitem{RHIC}
E.~V.~Shuryak,
\emph{What RHIC experiments and theory tell us about properties of  quark-gluon plasma?,}
\npa{750}{2005}{64}
[\hepph{0405066}].

\bibitem{RHIC-violates-1}
P.~Romatschke and U.~Romatschke,
\emph{Viscosity information from relativistic nuclear collisions: How perfect is the fluid observed at RHIC?,}
\prl{99}{2007}{172301}
[\arXivid{0706.1522} [nucl-th]].

\bibitem{RHIC-violates-2}
H.~Song and U.~W.~Heinz,
\emph{Suppression of elliptic flow in a minimally viscous quark-gluon plasma,}
\plb{658}{2008}{279}
[\arXivid{0709.0742} [nucl-th]].

\bibitem{swampland}
C.~Vafa,
\emph{The string landscape and the swampland,}
\hepth{0509212}.

\bibitem{weak-grav}
N.~Arkani-Hamed, L.~Motl, A.~Nicolis and C.~Vafa,
\emph{The string landscape, black holes and gravity as the weakest force,}
\jhep{06}{2007}{060}
[\hepth{0601001}].

\bibitem{causality}
A.~Adams, N.~Arkani-Hamed, S.~Dubovsky, A.~Nicolis and R.~Rattazzi,
\emph{Causality, analyticity and an IR obstruction to UV completion,}
\jhep{10}{2006}{014}
[\hepth{0602178}].

\bibitem{o-vafa-swamp}
H.~Ooguri and C.~Vafa,
\emph{On the geometry of the string landscape and the swampland,}
\npb{766}{2007}{21}
[\hepth{0605264}].

\bibitem{ebh}
Y.~Kats, L.~Motl and M.~Padi,
\emph{Higher-order corrections to mass-charge relation of extremal black holes,}
\jhep{12}{2007}{068}
[\hepth{0606100}].

\bibitem{Blau-Narain-Gava}
M.~Blau, K.~S.~Narain and E.~Gava,
\emph{On subleading contributions to the AdS/CFT trace anomaly,}
\jhep{09}{1999}{018}
[\hepth{9904179}].

\bibitem{Nojiri-Odintsov}
S.~Nojiri and S.~D.~Odintsov,
\emph{On the conformal anomaly from higher derivative gravity in AdS/CFT correspondence,}
\ijmpa{15}{2000}{413}
[\hepth{9903033}].

\bibitem{Henningson-Skenderis}
M.~Henningson and K.~Skenderis,
\emph{The holographic Weyl anomaly,}
\jhep{07}{1998}{023}
[\hepth{9806087}].

\bibitem{Gubser}
S.~S.~Gubser,
\emph{Einstein manifolds and conformal field theories,}
\prd{59}{1999}{025006}
[\hepth{9807164}].

\bibitem{Sp(N)}
A.~Fayyazuddin and M.~Spalinski,
\emph{Large $N$ superconformal gauge theories and supergravity orientifolds,}
\npb{535}{1998}{219}
[\hepth{9805096}].

\bibitem{AFM}
O.~Aharony, A.~Fayyazuddin and J.~M.~Maldacena,
\emph{The large $N$ limit of $\c{N} = 2,1$ field theories from three-branes in F-theory,}
\jhep{07}{1998}{013}
[\hepth{9806159}].

\bibitem{Aharony}
O.~Aharony, J.~Pawelczyk, S.~Theisen and S.~Yankielowicz,
\emph{A note on anomalies in the AdS/CFT correspondence,}
\prd{60}{1999}{066001}
[\hepth{9901134}].

\bibitem{justification}
A.~Buchel, R.~C.~Myers and A.~Sinha,
\emph{Beyond $\eta/s = 1/4\pi$,}
\arXivid{0812.2521} [hep-th].

\bibitem{Unruh}
W.~G.~Unruh,
\emph{Notes on black hole evaporation,}
\prd{14}{1976}{870}.

\bibitem{Wald-1}
R.~M.~Wald,
\emph{Black hole entropy is the Noether charge,}
\prd{48}{1993}{3427}
[\grqc{9307038}].

\bibitem{Wald-2}
V.~Iyer and R.~M.~Wald,
\emph{Some properties of Noether charge and a proposal for dynamical black hole entropy,}
\prd{50}{1994}{846}
[\grqc{9403028}].

\bibitem{Jacobson-Kang-Myers}
T.~Jacobson, G.~Kang and R.~C.~Myers,
\emph{On black hole entropy,}
\prd{49}{1994}{6587}
[\grqc{9312023}].

\bibitem{Brustein-Gorbonos-Hadad}
R.~Brustein, D.~Gorbonos and M.~Hadad,
\emph{Wald's entropy is equal to a quarter of the horizon area in units of the effective gravitational coupling,}
\arXivid{0712.3206} [hep-th].

\bibitem{Shenker}
M.~Brigante, H.~Liu, R.~C.~Meyers, S.~Shenker and S.~Yaida,
\emph{Viscosity bound violation in higher derivative gravity,}
\prd{77}{2008}{126006}
[\arXivid{0712.0805} [hep-th]].

\bibitem{RN-corr}
M.~Campanelli, C.~O.~Lousto and J.~Audretsch,
\emph{Perturbative method to solve fourth-order gravity field equations,}
\prd{49}{1994}{5188} [\grqc{9401013}].

\vspace{-4mm}

\end{thebibliography}
\end{document}